\documentclass[prb,twocolumn,showpacs,amsmath,amssymb,superscriptaddress]{revtex4}

\usepackage[dvipdfmx]{graphicx}
\usepackage{pdfpages}
\usepackage{dcolumn}
\usepackage{bm}
\usepackage{color}
\usepackage{amssymb}

\begin{document}

\title{
Ground-state properties of the $K-\Gamma$ model on a honeycomb lattice
}

\author{Takuto Yamada}
\author{Takafumi Suzuki}
\author{ Sei-ichiro Suga}

\affiliation{Graduate School of Engineering, University of Hyogo, Himeji 671-2280, Japan}

\date{\today}

\begin{abstract} 
We investigate the ground-state properies of the $K-\Gamma$ model on a honeycomb lattice using series expansions and numerical exact diagonalizations, where the model includes Kitaev ($K$) and symmetric off-diagonal ($\Gamma$) interactions. 
Starting from the weakly interacting dimers on the specific bond, we strengthen the interdimer interactions to the isotropically interacting system. We show that depending on $\Gamma$ and $K$, the dimer state survives up to the isotropically interacting system, where the phase transition occurs, or obeys a phase transition to a magnetically ordered state at an anisotropic interaction. The results are summarized in the phase diagram. We also show that the Kekul\'{e} dimerized state is unstable in the isotropic $K-\Gamma$ model. 
\end{abstract}

\preprint{APS/123-QED}

\maketitle

\section{\label{sec:level1}Introduction}
The realization of a spin liquid state has been a central issue in condensed matter physics.
It has been shown exactly that the ground state of the Kitaev model on a honeycomb lattice is a spin liquid, i.e., Kitaev spin liquid (KSL) \cite{Kitaev}. Two types of Majorana fermions emerge owing to the fractionalization of $S=1/2$ quantum spins \cite{Kitaev}. In relation to the appearance of Majorana fermions, exotic features have been predicted in thermodynamic quantities, spin dynamics, and transport properties \cite{Kitaev,Baskaran,Knolle,Nasu1}. 
There are several candidate materials, in which the Kitaev interaction is realized \cite{Khaliullin,Trebst,WinterRev,Takagi}. Specifically, $\alpha$-RuCl$_3$ is a promising candidate material that supports the Kitaev interaction between $j_{\rm eff}=1/2$ pseudospin moments  \cite{Plump,Kubota,Sandilands,Do,DHirobe,Sears,Banerjee1,Banerjee2,Banerjee3}. 
To discuss the experimental features of $\alpha$-RuCl$_3$, effective models have been proposed, which include not only the Kitaev interaction but also other interactions such as the Heisenberg interaction, symmetric off-diagonal interaction, and/or their further-neighbor interactions. The strength of the interactions has been evaluated using {\it ab-initio} calculations and {\it ab-initio}-guided empirical approaches.
According to these studies, it has been argued that the nearest-neighbor Kitaev interaction is the strongest, and the nearest-neighbor symmetric off-diagonal interaction is the second strongest \cite{Kim,Winter2016,Yadav,Winter2017,Suzuki}. In particular, effective models where these two interactions are dominant have succeeded in reproducing the key experimental features for the fractionalization of quantum spins qualitatively \cite{Kim} and quantitatively\cite{Suzuki}.  
These results suggest that the effective model that consists of the nearest-neighbor Kitaev ($K$) and symmetric off-diagonal  ($\Gamma$) interactions on a honeycomb lattice is a minimum model for $\alpha$-RuCl$_3$. 
This effective model is called the $K-\Gamma$ model \cite{Yamaji2018-1,Yamaji2018-2,AMSamarakoon,JWang,JWang2}.

Aside from the adequate description of $\alpha$-RuCl$_3$, the $K-\Gamma$ model itself possesses fascinating properties. From numerical exact diagonalizations (ED) of a 24-site cluster, it has been shown that the ground-state phase diagram includes various states such as magnetically ordered states and KSL \cite{Rau}. 
In prticular, it has been pointed out that a quantum spin liquid (QSL) state appears close to $K=0$ \cite{Yamaji2018-1,Yamaji2018-2}. When the $K$ interaction on one bond is slightly stronger than the $K$ interactions on the other two bonds, a first-order phase transition occurs between QSL and KSL \cite{Yamaji2018-1}. However, when the anisotropy of the $K$ interaction becomes stronger, QSL is adiabatically connected to KSL \cite{Yamaji2018-1}. 
The {\it ab-initio} calculations for $\alpha$-RuCl$_3$ have also predicted that $K$ and $\Gamma$ interactions on one of the three bonds are stronger than those on the other two bonds \cite{Kim}. These findings motivate us to investigate the ground-state properties of the $K-\Gamma$ model by including the anisotropy of the interactions.

In this study, we investigate the ground-state phase diagram of the $S=1/2$ $K-\Gamma$ model on a honeycomb lattice by employing series expansions \cite{textbook,JOitmaa}. Series expansions are based on graph theories and can systematically include higher order terms. Thus, series expansions are used complementarily with ED. 
We adopt dimer series expansions, in which the interactions on specific bonds are included in the initial state, and the interactions between the dimers on specific bonds are included perturbatively. 
We calculate the ground-state energy and its first and second derivatives. We also calculate them with 24-site ED. By combining the results obtained using these two methods, we investigate the stable ground state when the interactions between the dimers become strong towards the isotropically interacting system.

In Sec. \ref{sec:level2}, we introduce the model and outline dimer series expansions. 
In Sec. \ref{sec:level3}, we explain the details of dimer series expansions and show the numerical results. We discuss the stable state when the interdimer interaction becomes strong and the system changes from the anisotropically interacting dimer system to the isotropically interacting system for given $K$ and $\Gamma$ interactions. By summarizing the results, we obtain the phase diagram. We show that depending on $K$ and $\Gamma$, the dimer state obeys a phase transition at an anisotropic interaction or survives up to the isotropically interacting system where a phase transition occurs. We also propose two scenarios for the connection of QSL \cite{Yamaji2018-1,Yamaji2018-2} to the dimer state. 
In Sec. \ref{sec:level4}, we discuss a stable ordered state that has discrete symmetry at the isotropically interacting system. For this purpose, we adopt the ${\mathcal T}_6$ transformation \cite{JChaloupkaPRB} to the $K-\Gamma$ model. We then perform dimer series expansions to the transformed model that has a Kekul\'e structure concerning the interaction. A summary is provided in Sec. \ref{sec:level5}.

\section{\label{sec:level2}Model and Method}
\begin{figure}[htb]
\begin{center}
\includegraphics[width=0.95\hsize]{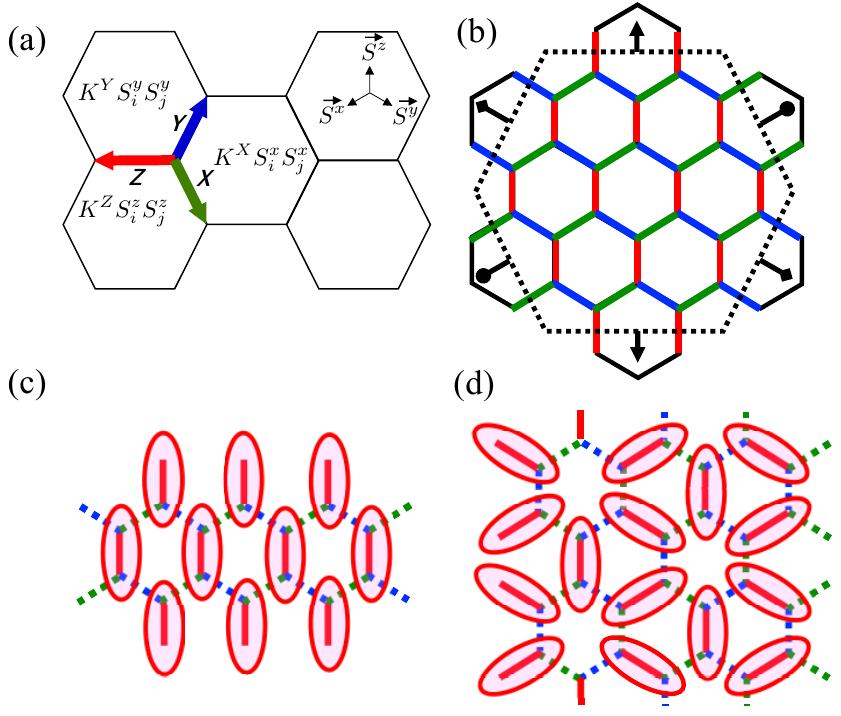}
\hspace{0pc}
\vspace{0pc}
\caption{(Color online) \label{fig1} (a) Kitaev-type interactions on a honeycomb lattice. 
(b) $24$-site cluster. 
Periodic boundary conditions are applied on the dotted lines with common symbols.
(c) Schematic picture of dimer series expansions performed in Sec. \ref{sec:level3}. Ellipsoids denote the initial spin dimers on $Z$ bonds. Green and blue dotted lines denote the interactions on $X$ and $Y$ bonds, which have the same strengths.  
(d) Schematic picture of dimer series expansions performed in Sec. \ref{sec:level4}. The ${\mathcal T}_6$ transformation \cite{JChaloupkaPRB} changes not only the geometric pattern of $X$, $Y$, and $Z$ bonds but also the interaction. The interaction on each bond is transformed into the $XXZ$-type one, $-K{S_i}^{\gamma}{S_j}^{\gamma}-\Gamma ({S_i}^{\alpha}{S_j}^{\alpha}+{S_i}^{\beta}{S_j}^{\beta})$.
Ellipsoids denote the initial spin dimers. 
}
\end{center}
\end{figure}
The $S=1/2$ honeycomb $K-\Gamma$ model is defined as
\begin{eqnarray}
{\mathcal H}= \sum_{{\langle ij \rangle}_{\gamma=X,Y,Z}}  \left[ K {S_i}^{\gamma} {S_j}^{\gamma} + \Gamma \left( {S_i}^{\alpha} {S_j}^{\beta} + {S_i}^{\beta} {S_j}^{\alpha} \right)  \right], 
\end{eqnarray}
where $\langle ij \rangle_{\gamma = X,Y,Z}$ denotes the nearest-neighbor pair on the $\gamma$ bond of the honeycomb lattice, as shown in Fig. \ref{fig1}(a), and $\alpha$ and $\beta$ are the different spin components from the $\gamma$ component. 
We set $K=-\cos \theta$ and $\Gamma=\sin \theta, \;(0^{\circ} \leq \theta < 360^{\circ})$. Thus, $\theta=0^{\circ}$ ($180^{\circ}$) describes the Kitaev model with a ferromagnetic (FM) [antiferromagnetic (AF)] interaction, while $\theta=90^{\circ}$ ($\theta=270^{\circ}$) describes the $\Gamma$ model, which has only the AF   (FM) $\Gamma$ interaction.

To adopt dimer series expansions \cite{textbook}, we divide Hamiltonian (1) into the following perturbative form,  
\begin{eqnarray}
{\mathcal H} = {\mathcal H_{\rm D}}^0 + \lambda {\mathcal H_{\rm D}}^1, 
\end{eqnarray}
where ${\mathcal H}_{\rm D}^0$ describes the unperturbed initial dimers on $\gamma$ bonds, and ${\mathcal H}_{\rm D}^1$ describes the perturbative term with $0 \le \lambda \le 1$.

Starting from the isolated initial dimers on $\gamma$ bonds, we perform dimer series expansions with respect to $\lambda {\mathcal H_{\rm D}}^1$ up to the eighth order. 
We calculate the ground-state energy per unit cell, $E$, and its first and second derivatives, $\partial E/\partial \lambda$ and $\partial^2 E/\partial \lambda^2$. 
We also calculate these quantities with 24-site ED. The 24-site cluster is shown in Fig. \ref{fig1}(b).  
We discuss the ground-state property towards the isotropic point at $\lambda=1$.

\section{\label{sec:level3} Results for dimer series expansions }
We set the initial dimers on the $Z$ bonds, as shown in Fig. \ref{fig1}(c), and investigate the stability of such a dimer state towards the $\lambda=1$ isotropic system. Hamiltonian (2) reads   
\begin{eqnarray}
{\mathcal H_{\rm D}}^0 = \sum_{{\langle ij \rangle}_{Z}}  \left[ K {S_i}^{z} {S_j}^{z} + \Gamma \left( {S_i}^{x} {S_j}^{y} + {S_i}^{y} {S_j}^{x} \right)  \right],  \\
{\mathcal H_{\rm D}}^1 = \sum_{{\langle ij \rangle}_{\gamma=X,Y}}  \left[ K {S_i}^{\gamma} {S_j}^{\gamma} +  \Gamma \left( {S_i}^{\alpha} {S_j}^{\beta} + {S_i}^{\beta} {S_j}^{\alpha} \right)  \right].    
\end{eqnarray}
This situation is schematically shown in Fig. \ref{fig1}(c).
For the initial state, we have four candidates: singlet dimer $|s\rangle=(|\uparrow\downarrow\rangle - |\downarrow\uparrow\rangle)/\sqrt{2}$,  triplet dimers $|t_0\rangle=(|\uparrow\downarrow\rangle + |\downarrow\uparrow\rangle)/\sqrt{2}$, $|t_x\rangle=(|\uparrow\uparrow\rangle - i|\downarrow\downarrow\rangle)/\sqrt{2}$,   
and  $|t_y\rangle=(|\uparrow\uparrow\rangle + i|\downarrow\downarrow\rangle)/\sqrt{2}$, where the up and down arrows denote the spin up and down states, respectively.  
Depending on $\theta$, we adopt one of them that possesses the lowest energy. 
For the initial state, the parameter space is specified into three regions: At $0^{\circ} < \theta < 135^{\circ}$, the $|t_x\rangle$ dimer is the initial state; at $225^{\circ} < \theta < 360^{\circ}$, the $|t_y\rangle$ dimer is the initial state. At $135^{\circ} < \theta < 225^{\circ}$, the $|s\rangle$ and $|t_0\rangle$ dimers are degenerate, which makes it difficult to perform dimer series expansions.

\subsection{$\Gamma>0$}
\begin{figure}[htb]
\begin{center}
\includegraphics[width=0.45\hsize]{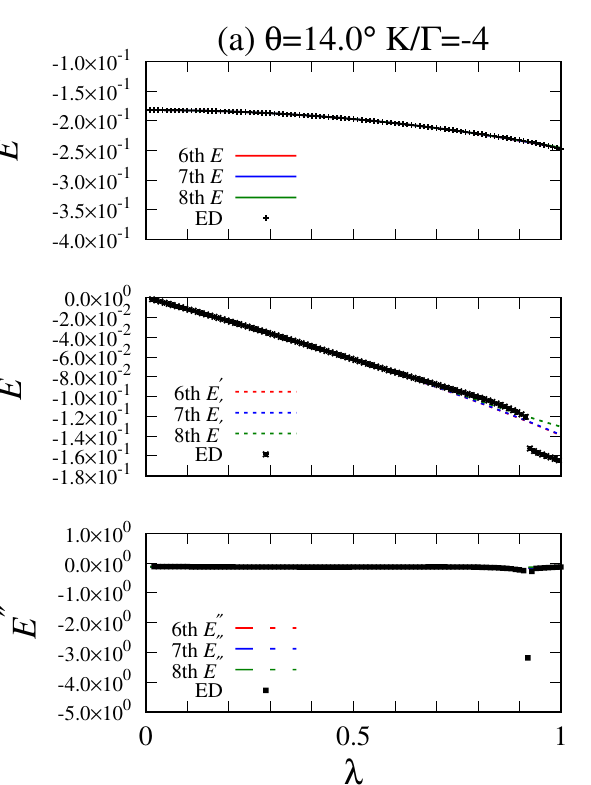}
\includegraphics[width=0.45\hsize]{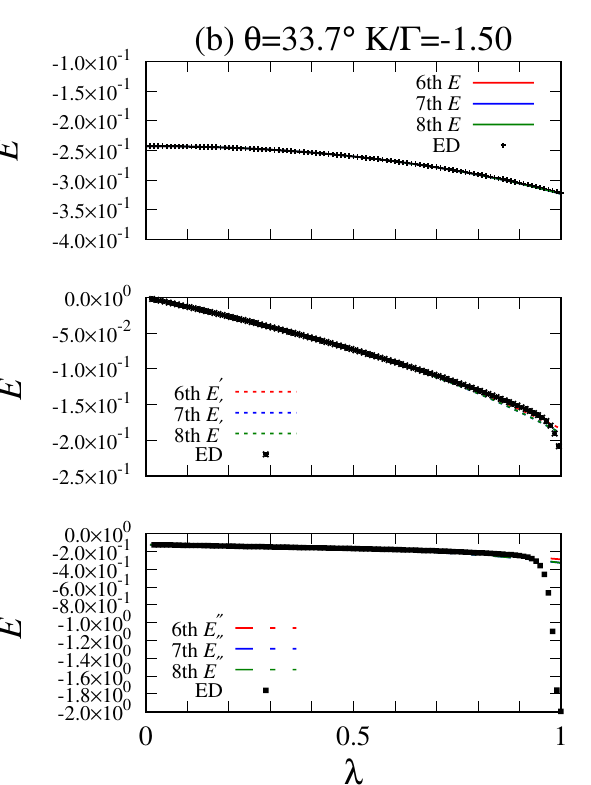}
\includegraphics[width=0.45\hsize]{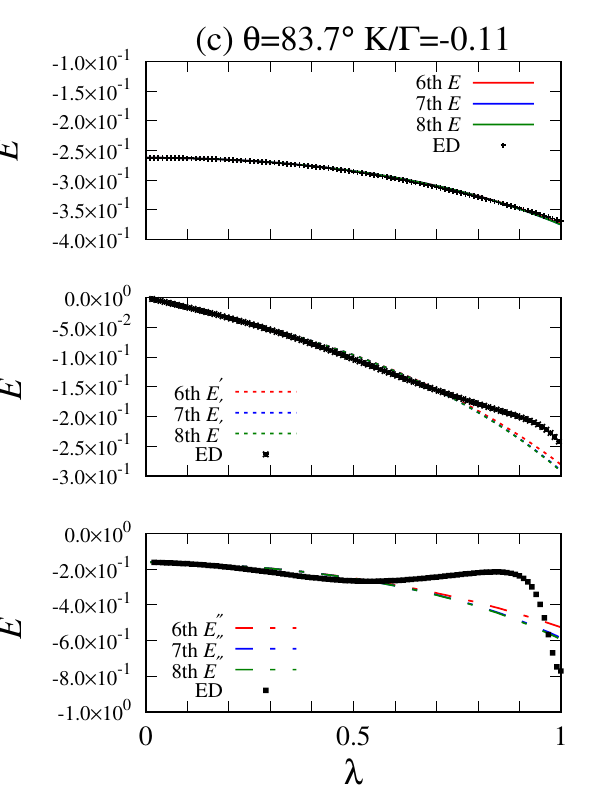}
\includegraphics[width=0.45\hsize]{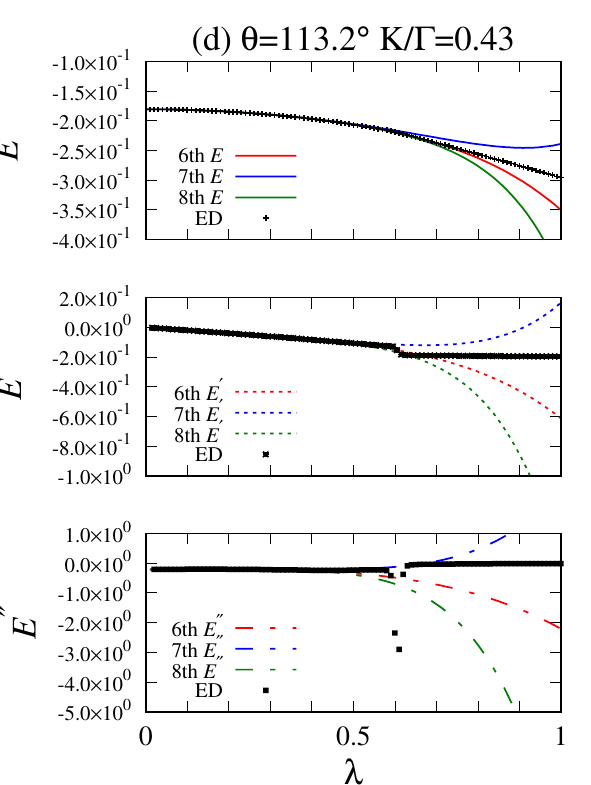}
\caption{(Color online) \label{fig2} For $\Gamma>0$, the typical behavior of $E$, $\partial E/\partial \lambda$, and $\partial^2 E/\partial \lambda^2$ obtained with dimer series expansions included up to sixth, seventh, and eighth orders and with 24-site ED. $\theta ={\rm(a)}\; 14.0^{\circ}$, ${\rm(b)}\; 33.7^{\circ}$, ${\rm(c)}\; 83.7^{\circ}$, and ${\rm(d)}\; 113.2^{\circ}$. 
}
\end{center}
\end{figure}
We first show $E$, $\partial E/\partial \lambda$, and $\partial^2 E/\partial \lambda^2$ for $\Gamma>0$ as a function of $\lambda$. 
Figures \ref{fig2}(a)--\ref{fig2}(d) show the typical behavior of these quantities up to sixth, seventh, and eighth orders for $\theta = 14.0^{\circ} \;(K/\Gamma=-4), 33.7^{\circ} \;(K/\Gamma=-1.50),  83.7^{\circ} \;(K/\Gamma=-0.11)$, and $113.2^{\circ} \;(K/\Gamma=0.43)$, respectively. 
We adopt the $|t_x\rangle$ dimer as the initial state.  
We compare these quantities with those obtained with 24-site ED.
For $\theta = 14.0^{\circ}, 33.7^{\circ}$, and $83.7^{\circ}$, the ground-state energies up to sixth, seventh, and eighth orders converge at $0 \leq \lambda \leq 1$. The ground-state energies agree with those obtained with ED. 
For $\theta = 14.0^{\circ}$, $\partial E/\partial \lambda$ obtained with ED changes discontinuously at $\lambda \approx 0.9$, where $\partial E/\partial \lambda$ up to sixth, seventh, and eighth orders begin to deviate with an increase in $\lambda$. 
The results mean that the $|t_x\rangle$-dimer state is stable at $0 \leq \lambda < 0.9$ and undergoes a first-order phase transition at  $\lambda \approx 0.9$. 
For $\theta=33.7^{\circ}$, $\partial^2 E/\partial \lambda^2$ obtained with ED steeply decreases towards $\lambda \approx 1$, where $\partial^2 E/\partial \lambda^2$ up to sixth, seventh, and eighth orders deviate from each other with an increase in $\lambda$. 
This behavior suggests that the $|t_x\rangle$-dimer state is stable at $0 \leq \lambda <1$ and undergoes a phase transition at $\lambda = 1$. 
For $\theta=83.7^{\circ}$, $\partial^2 E/\partial \lambda^2$ obtained using ED shows a small minimum at $\lambda \approx 0.55$ and steeply decreases towards $\lambda \approx 1$. 
At $\lambda \approx 0.55$, $\partial^2 E/\partial \lambda^2$ up to sixth, seventh, and eighth orders begin to slightly deviate with an increase in $\lambda$, which is caused by the lack of higher orders of the dimer series expansions.  
We consider that the small minimum at $\lambda \approx 0.55$ is caused by the finite-size effect or the vestige of a prominent minimum caused by a phase transition at $\theta>90^{\circ}$, i.e., crossover behavior. 
The results suggest that the $|t_x\rangle$-dimer state is stable at $0 \leq \lambda < 1$, and the isotropic point at $\lambda = 1$ is a phase transition point.

For $\theta=113.2^{\circ}$, $E$ up to sixth, seventh, and eighth orders begin to deviate from each other at $\lambda \approx 0.6$ with an increase in $\lambda$. The ground-state energy obtained with ED also deviates from them at $\lambda \approx 0.6$. 
The first derivative obtained with ED discontinuously changes at $\lambda \approx 0.6$, where $\partial E/\partial \lambda$ up to sixth, seventh, and eighth orders begin to deviate. At $\lambda \approx 0.6$, $\partial^2 E/\partial \lambda^2$ shows a dip.  
With an increase in the order of dimer series expansions, changes in $E$, $\partial E/\partial \lambda$, and $\partial^2 E/\partial \lambda^2$ towards $\lambda=1$ become clear, which suggests that dimer series expansions fail.   
The results mean that the $|t_x\rangle$-dimer state is stable at $0 \leq \lambda < 0.6$ and undergoes a second-order or a weak first-order phase transition at  $\lambda \approx 0.6$. 
This phase transition is related to the emergence of the $120^{\circ}$ magnetically ordered phase \cite{Rau,Yamaji2018-1,Yamaji2018-2}, because neither $\partial E/\partial \lambda$ nor $\partial^2 E/\partial \lambda^2$ obtained with ED shows a singularity at $\lambda>0.6$.

\subsection{$\Gamma<0$}
\begin{figure}[htb]
\begin{center}
\includegraphics[width=0.45\hsize]{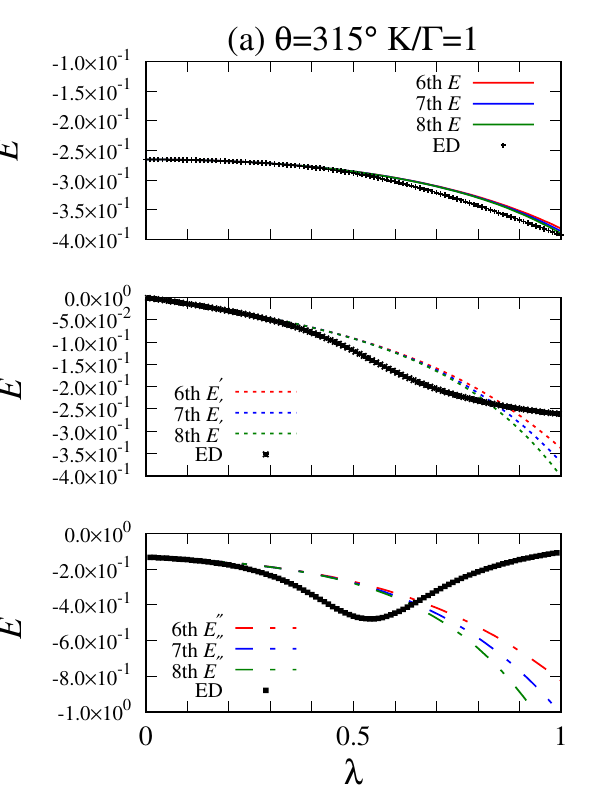}
\includegraphics[width=0.45\hsize]{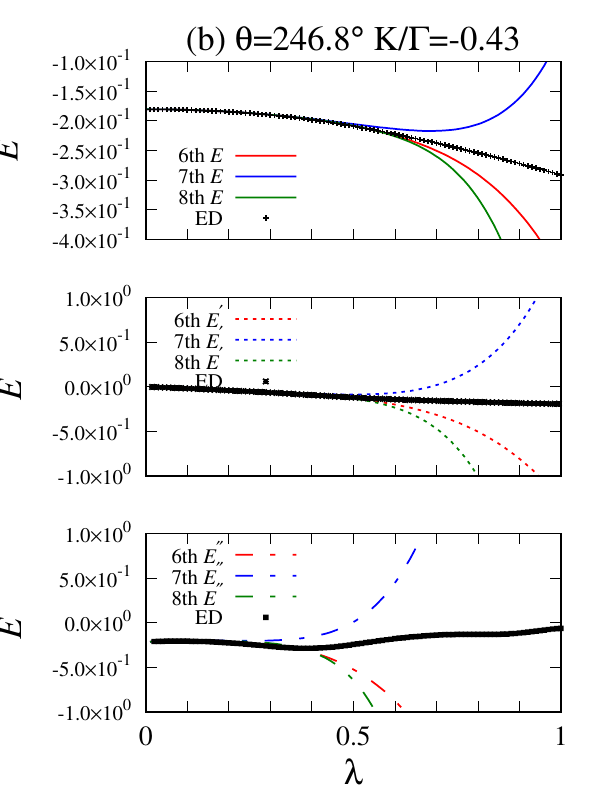}
\hspace{0pc}
\vspace{0pc}
\caption{(Color online) \label{fig3}  For $\Gamma<0$, the typical behavior of $E$, $\partial E/\partial \lambda$, and $\partial^2 E/\partial \lambda^2$ obtained with dimer series expansions included up to sixth, seventh, and eighth orders and with 24-site ED. $\theta={\rm(a)}\; 315^{\circ}$ and ${\rm(b)}\; 246.8^{\circ}$. 
}
\end{center}
\end{figure}
Figures \ref{fig3}(a) and \ref{fig3}(b) show the typical behavior of $E$, $\partial E/\partial \lambda$, and $\partial^2 E/\partial \lambda^2$ for $\Gamma<0$. We adopt the $|t_y\rangle$ dimer as the initial state.  
For  $\theta=315^{\circ} \;(K/\Gamma=1)$, $E$ up to sixth, seventh, and eighth orders begin to slightly deviate from $E$ obtained with ED at $\lambda \approx 0.55$ with an increase in $\lambda$, and $\partial^2 E/\partial \lambda^2$ up to sixth, seventh, and eighth orders also begin to deviate. 
The second derivative, $\partial^2 E/\partial \lambda^2$, obtained with ED shows a minimum at $\lambda \approx 0.55$, which is notable compared to that at $\lambda \approx 0.55$ for $\theta=83.7^{\circ}$ shown in Fig. 2(c).   
Thus, we consider that phase transition occurs from the $|t_y\rangle$-dimer state to the ferromagnetically ordered state \cite{Rau} at $\lambda \approx 0.55$. 
Compared to the results shown in Fig. 2(c), $\partial^2 E/\partial \lambda^2$ obtained with ED does not show a steep decrease towards $\lambda=1$ but shows a notable minimum at $\lambda \gg 1$ (not shown), where the system approaches a spin chain that consists of $K$ and $\Gamma$ interactions on $X$ and $Y$ bonds. At $\lambda \gg 1$, the $K$ and $\Gamma$ interactions on the $Z$ bond act as relevant interchain interactions to the spin chain, which leads to the phase transition to a magnetically ordered state.  
When $\lambda$ is decreased from $\lambda \gg 1$ to $\lambda \approx 0.55$, another minimum does not appear at $\partial^2 E/\partial \lambda^2$ obtained with ED. In addition, on the basis of the abovementioned results, we consider that the phase transition to a magnetically ordered state occurs at $\lambda \approx 0.55$.

For $\theta=246.8^{\circ} \;(K/\Gamma=-0.43)$,  $E$, $\partial E/\partial \lambda$, and $\partial^2 E/\partial \lambda^2$ up to sixth, seventh, and eighth orders start to deviate at $\lambda \approx 0.6$ with an increase in $\lambda$. 
These values increase and decrease towards $\lambda=1$ depending on the included order. 
Close to $\lambda=1$, their changes become more conspicuous when the higher order terms are included. 
The results mean that dimer series expansions fail in $\lambda \gtrapprox 0.6$.
The second derivative, $\partial^2 E/\partial \lambda^2$, obtained with ED shows the local minima at $\lambda \approx 0.39$ and $0.83$. 
Thus, we conclude that the $|t_y\rangle$-dimer state becomes unstable at $\lambda \gtrapprox 0.39$.

\subsection{Phase diagram }
\begin{figure}[htb]
\begin{center}
\includegraphics[width=0.45\hsize]{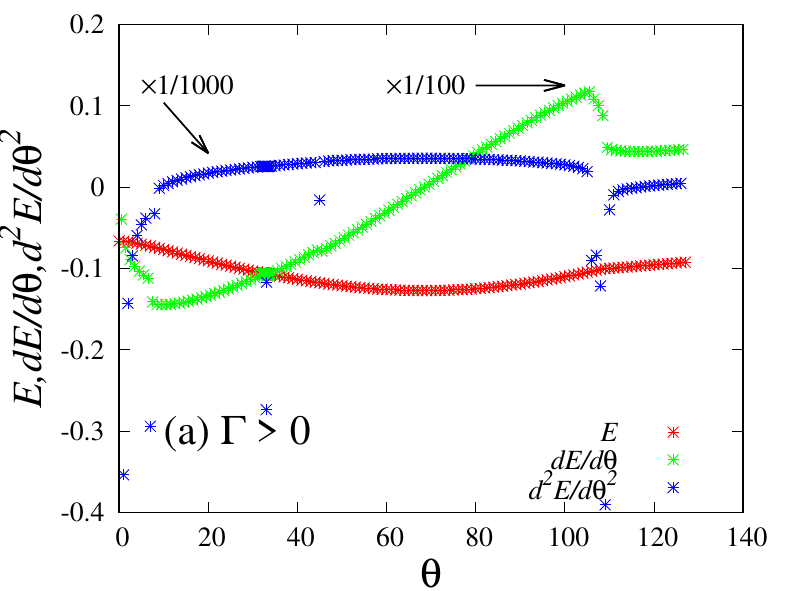}
\includegraphics[width=0.45\hsize]{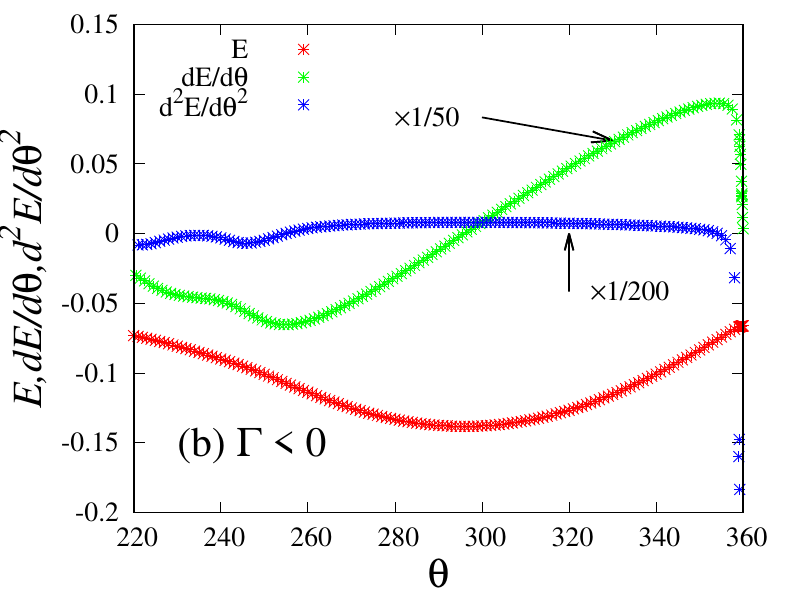}
\includegraphics[width=0.45\hsize]{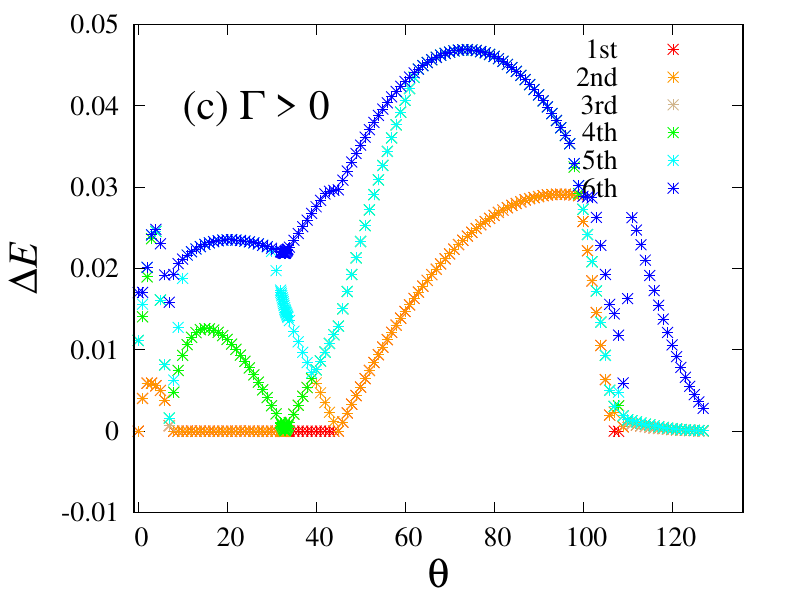}
\includegraphics[width=0.45\hsize]{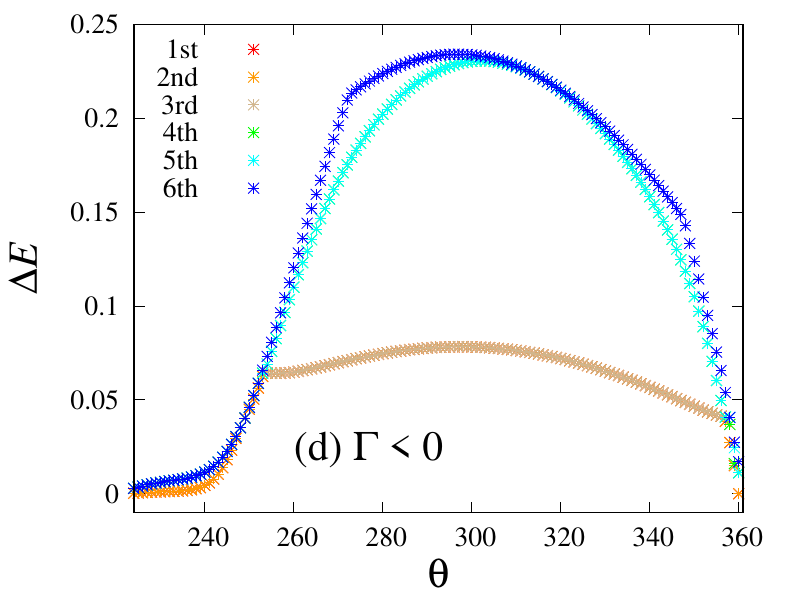}
\hspace{0pc}
\vspace{0pc}
\caption{(Color online) \label{fig4} $E$, $\partial E/\partial \theta$, and $\partial^2 E/\partial \theta^2$ at $\lambda=1$ obtained with 24-site ED for (a) $\Gamma>0$ and (b) $\Gamma<0$. Excitation energies at $\lambda=1$ obtained with 24-site ED for (c) $\Gamma>0$ and (d) $\Gamma<0$. The ordinal numbers denote the excited states.  
}
\end{center}
\end{figure}
\begin{figure}[htb]
\begin{center}
\includegraphics[width=0.7\hsize]{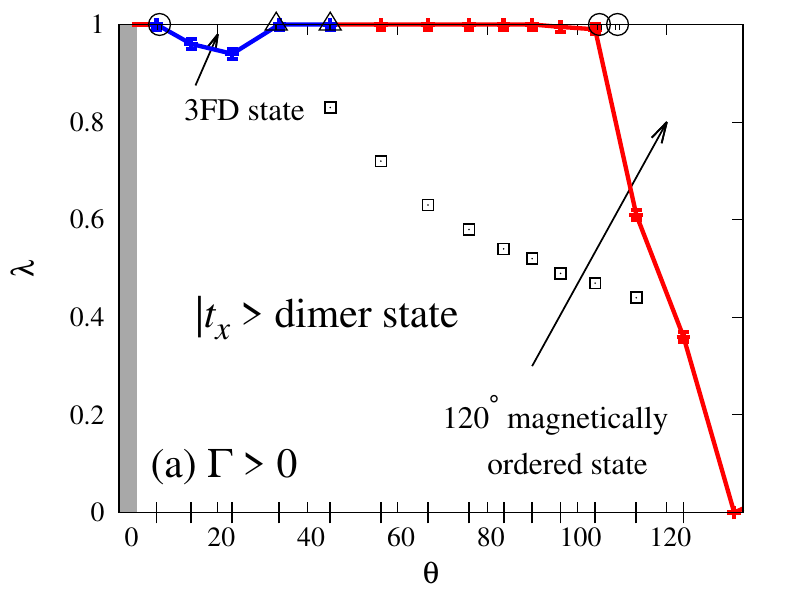}
\includegraphics[width=0.7\hsize]{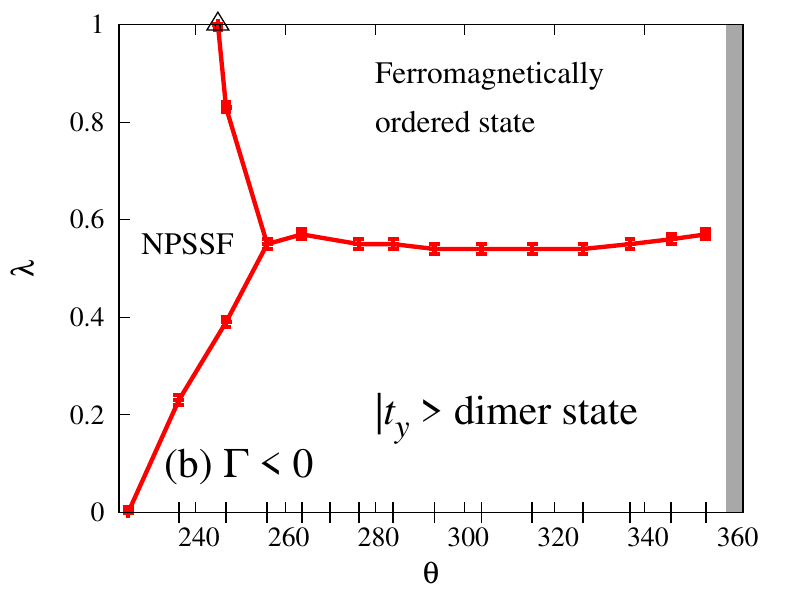}
\hspace{0pc}
\vspace{0pc}
\caption{(Color online) \label{fig5} Phase diagrams for (a) $\Gamma>0$ and (b) $\Gamma<0$. 
At (a) ``3 FD state'', this state is threefold degenerate (3 FD). At (b) ``NPSSF'', we find no prominent peak in the static spin structure factor (NPSSF).
Symbols for the phase boundaries are determines from $\partial E/\partial \lambda$ and $\partial^2 E/\partial \lambda^2$ calculated with 24-site ED.   
Open circle means a jump or a cusp in $\partial E/\partial \theta$, which indicates a first-order phase transition. 
Open triangle means a clear minimum in $\partial^2 E/\partial \theta^2$, which leads to a second-order phase transition or a weak first-order phase transition in the thermodynamic limit. 
Blue cross indicates the first-order phase transition point, while the red cross indicates the second-order or weak first-order phase transition. 
Open square means a crossover point where $\partial^2 E/\partial \lambda^2$ shows a small minimum.
Error bars are smaller than the symbol sizes. Lines are guide for the eye. 
KSL phases, which are expected to appear in the gray hatched area, are difficult to determine because of the numerical resolution.}
\end{center}
\end{figure}
To determine the phase boundaries at $\lambda=1$, we calculate $\partial E/\partial \theta$ and $\partial^2 E/\partial \theta^2$ with 24-site ED and show the results for $\Gamma>0$ and $\Gamma<0$ in Figs. \ref{fig4}(a) and \ref{fig4}(b), respectively. 
The discontinuity or cusp in $\partial E/\partial \theta$ appears at $\theta=7^{\circ} \pm 1^{\circ} \;(K/\Gamma=-8.14 \pm 1.37)$, $105^{\circ} \pm 1^{\circ} \;(K/\Gamma=0.27 \pm 0.02)$, and $109^{\circ} \pm 1^{\circ} \;(K/\Gamma=0.34 \pm 0.02)$ for $\Gamma>0$, which indicates that the first-order phase transitions in terms of $\theta$ occur there. The second derivative, $\partial^2 E/\partial \theta^2$, for $\Gamma>0$ shows dips at $\theta=33.2^{\circ} \pm 0.1^{\circ} \;(K/\Gamma=-1.53 \pm 0.01)$ and $\theta=45.0^{\circ} \pm 0.2^{\circ} \;(K/\Gamma=-1\pm 0.01)$. The results indicate that a weak first-order transition or a second-order transition occurs there. For $\Gamma<0$, the minimum in $\partial^2 E/\partial \theta^2$ appears at $\theta=255^{\circ}\pm 1^{\circ} \;(K/\Gamma=-0.27 \pm 0.02)$, which corresponds to the phase transition to the magnetically ordered state \cite{Rau}.
Although KSL is expected to appear close to $\theta \approx 0^{\circ}$ and $360^{\circ}$, the signs of the KSL phase boundaries are difficult to determine because of the computational resolution.

Next, we calculate excitation energies at $\lambda=1$ with 24-site ED \cite{HPhi}, and show the results for $\Gamma>0$ and $\Gamma<0$ in Figs. \ref{fig4}(c) and \ref{fig4}(d), respectively.  
At $7^{\circ} \pm 1^{\circ}<\theta <33.2^{\circ} \pm 0.1^{\circ}$ the ground state is threefold degenerate. According to the phase diagram of Fig. 3(a) in Ref. \onlinecite{Rau}, this state is in the spiral phase. On the other hand, according to the phase diagram of Fig. 4(a) in Ref. \onlinecite{Kaneko}, this state is in the 6-site order phase with threefold degeneracy. The nature of the state in this parameter region is still under debate. We have shown that the ground state of this parameter region at $\lambda=1$ is threefold degenerate and, as will be shown later, the threefold degenerate state extends to the anisotropic $\lambda < 1$ region.  
At $33.2^{\circ} \pm 0.1^{\circ}<\theta < 45.0^{\circ}\pm{0.2}^{\circ}$ the ground state is twofold degenerate.  
However, at $0^{\circ} \lessapprox \theta <7^{\circ} \pm 1^{\circ}$ and $45.0^{\circ}\pm{0.2}^{\circ}<\theta<105^{\circ}\pm 1^{\circ}$, the ground state is unique in the 24-site cluster. 
At $109^{\circ} \pm 1^{\circ}<\theta<135^{\circ}$, the ground state is sixfold degenerate, which reflects the $120^{\circ}$ magnetically ordered state \cite{Rau,JRausnacko}.
At $225^{\circ} <\theta < 255^{\circ}\pm1^{\circ}$, the spiral phase is expected \cite{Rau}. 
The unique ground state at $255^{\circ}\pm1^{\circ}<\theta<360^{\circ}$ suggests the ferromagnetically ordered state \cite{Rau}.

We calculate $E$, $\partial E/\partial \lambda$, and $\partial^2 E/\partial \lambda^2$ by systematically changing $\theta$.  
Thus far, the results for $\Gamma>0$ and $\Gamma<0$ are summarized in the phase diagrams shown in Figs. \ref{fig5}(a) and \ref{fig5}(b), respectively.
At $0^{\circ} \lessapprox \theta < 7^{\circ} \pm 1^{\circ}$ and $33.2^{\circ} \pm 0.1^{\circ} < \theta < 105^{\circ} \pm 1^{\circ}$, the $|t_x\rangle$-dimer state is stable at $0\leq \lambda<1$, and the isotropic point ($\lambda =1$) is a phase transition point.
At $7^{\circ}\pm1^{\circ} < \theta < 33.2^{\circ}\pm 0.1^{\circ}$, the first-order phase transition occurs from the $|t_x\rangle$-dimer state to the threefold degenerate state with an increase in $\lambda$. 
At $109^{\circ}\pm 1^{\circ}<\theta$ the results suggest that a phase transition to the $120^{\circ}$ magnetically ordered state occurs \cite{Yamaji2018-2,Rau,JRausnacko}. 
At $135^{\circ}<\theta$ for $\Gamma>0$, dimer series expansions fail. 
Figure \ref{fig5}(b) indicates that at $225^{\circ}\pm1^{\circ}<\theta <360^{\circ}$, the $|t_y\rangle$-dimer state is stable at $0 \leq \lambda < 0.55$, while the ferromagnetically ordered state is stable at $\lambda>0.55$. 
At $\theta \leq 255^{\circ}\pm1^{\circ}$ for $\Gamma<0$, the $|t_y\rangle$-dimer state undergoes a phase transition to the threefold degenerate state. The $|t_y\rangle$-dimer region is reduced with a decrease in $\theta$ and disappears at $\theta=225^{\circ}$. In the state for larger $\lambda$ than the $|t_y\rangle$-dimer state, no prominent peak appears in the static spin structure factor up to $\lambda=1$ (not shown). To elucidate the characteristics, we have to calculate a larger-size system.

The $|t_x\rangle$-dimer ($|t_y\rangle$-dimer) state at $0< \theta<90^{\circ}$ ($225^{\circ}< \theta<270^{\circ}$) shows various features compared with the $|t_x\rangle$-dimer ($|t_y\rangle$-dimer) state at $90< \theta<135^{\circ}$ ($270^{\circ}< \theta<360^{\circ}$). These differences are considered to be attributed to the additional frustration effects between $K$ and $\Gamma$ with opposite signs. When we consider the strong $\Gamma$ limit, a spin configuration can be fixed depending on the sign of $\Gamma$. At $0^{\circ} < \theta < 90^{\circ}$ and $225^{\circ}< \theta<270^{\circ}$, the signs of $K$ and $\Gamma$ are opposite, which leads to the additional frustration effects against the spin configuration in the strong $\Gamma$ limit. At $90< \theta<135^{\circ}$ and $270^{\circ} < \theta < 360^{\circ}$, the signs of $K$ and $\Gamma$ are the same and, thus, no such additional frustration emerges.

\begin{figure}[htb]
\begin{center}
\includegraphics[width=0.45\hsize]{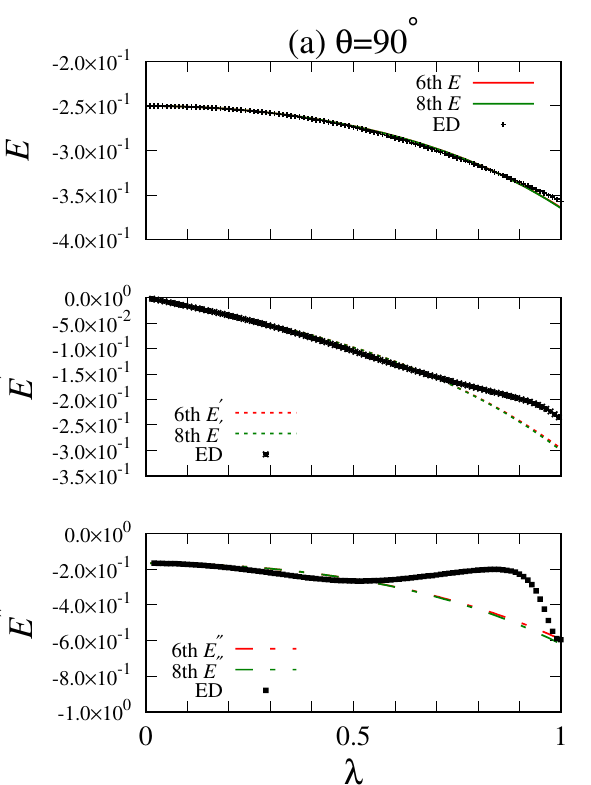}
\includegraphics[width=0.45\hsize]{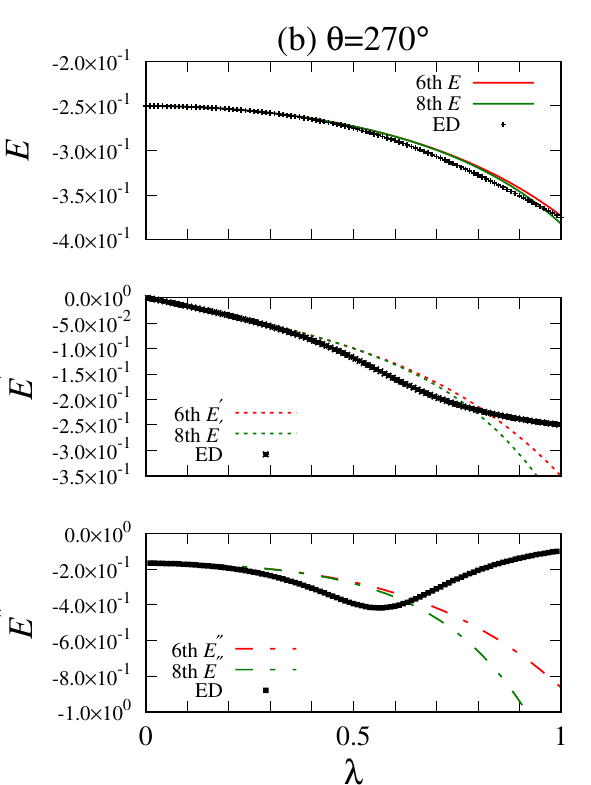}
\includegraphics[width=0.45\hsize]{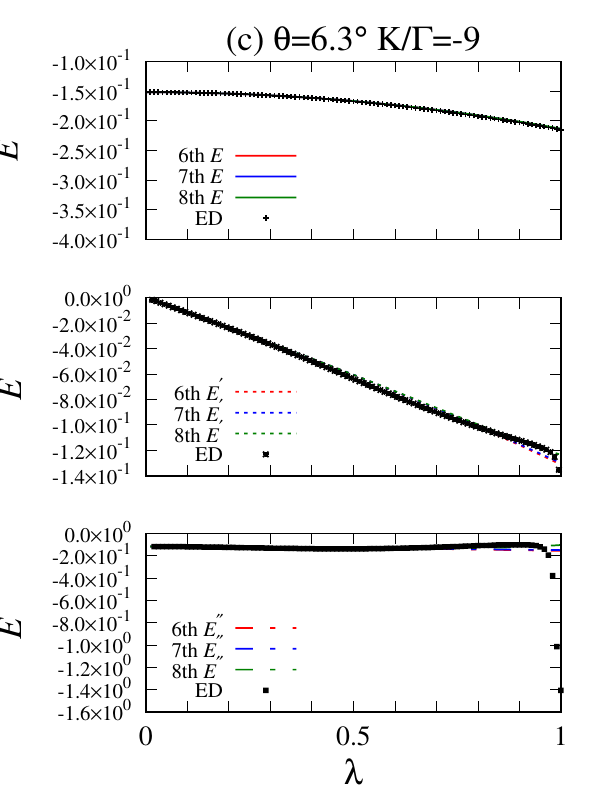}
\includegraphics[width=0.45\hsize]{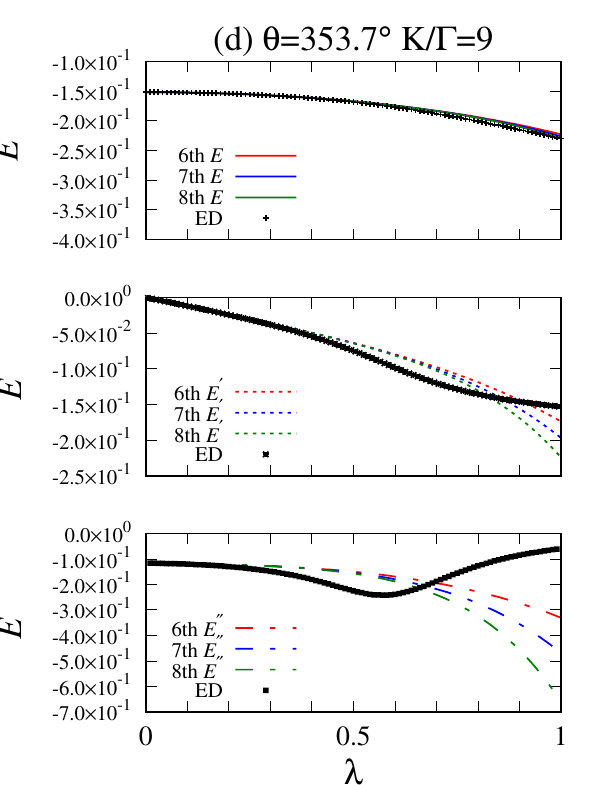}
\hspace{0pc}
\vspace{0pc}
\caption{(Color online) \label{fig6} The typical behavior of $E$, $\partial E/\partial \lambda$, and $\partial^2 E/\partial \lambda^2$ obtained with dimer series expansions up to sixth, seventh, and eighth orders and with 24-site ED. 
  $\theta={\rm(a)}\; 90^{\circ}$, ${\rm(b)}\; 270^{\circ}$, ${\rm(c)}\; 6.3^{\circ}$, and ${\rm(d)}\; 353.7^{\circ}$. }
\end{center}
\end{figure}
We now investigate the properties for $\theta=90^{\circ}$ and $270^{\circ}$, where the AF and FM isotropic $\Gamma$ models are realized at $\lambda=1$, respectively. It has been argued that a QSL state appears close to  $K=0$ in the isotropic and anisotropic $K-\Gamma$ models \cite{Yamaji2018-1,Yamaji2018-2}.  
Figures \ref{fig6}(a) and \ref{fig6}(b) show $E$, $\partial E/\partial \lambda$, and $\partial^2 E/\partial \lambda^2$ for $\theta=90^{\circ}$ and $270^{\circ}$, respectively. 
Note that at $K=0$, all coefficients of the odd order in the dimer series expansions become zero because of cancellation. 
We compare them with those for $\theta=6.3^{\circ} \;(K/\Gamma=-9.06)$ [Fig. \ref{fig6}(c)] and $353.7^{\circ} \;(K/\Gamma=9.06)$ [Fig. \ref{fig6}(d)], where the systems possess AF and FM $\Gamma$ interactions, respectively, and are located close to or on the KSL phase caused by the FM Kitaev interaction.

In these four states, the ground-state energies obtained with the dimer series expansions converge and agree with those obtained with ED. The first derivatives do not show discontinuity. A difference appears in $\partial^2 E/\partial \lambda^2$.  
For $\theta=90^{\circ}$, $\partial^2 E/\partial \lambda^2$ obtained with ED shows a small minimum at $\lambda \approx 0.55$, while for  $\theta=6.3^{\circ}$, $\partial^2 E/\partial \lambda^2$ shows a steep decrease towards $\lambda=1$. 
For $\theta=90^{\circ}$, $\partial^2 E/\partial \lambda^2$ obtained with dimer series expansions also decrease towards $\lambda=1$, and their decrease becomes slightly pronounced with an increase in the included order. However, for $\theta=6.3^{\circ}$, $\partial^2 E/\partial \lambda^2$ obtained with dimer series expansions do not show such notable decrease towards $\lambda=1$. 
The results indicate that at $\lambda=1$, the AF $\Gamma$ model and the $K-\Gamma$ model at $\theta=6.3^{\circ}$ are on a phase transition point. 
At least, the $|t_x\rangle$-dimer state survives in the vicinity of $\lambda=1$ for $\theta=6.3^{\circ}$. For $\theta=90^{\circ}$, the obtained results allow us to propose alternative scenarios. (i) The $|t_x\rangle$-dimer state  survives up to $\lambda=1$. (ii) The $|t_x\rangle$-dimer state becomes unstable at $\lambda \approx 0.55$, and another state obeys a phase transition at $\lambda=1$.

The second derivatives for $\theta=270^{\circ}$ and $353.7^{\circ}$ qualitatively show similar behavior. Specifically, $\partial^2 E/\partial \lambda^2$ obtained with dimer series expansions decrease towards $\lambda=1$ and their decrease becomes pronounced with an increase in the included order.  
However, $\partial^2 E/\partial \lambda^2$ obtained with ED does not show a decrease close to $\lambda=1$ but shows a minimum at $\lambda \approx 0.55$. The results suggest that the $|t_y\rangle$-dimer state becomes unstable at $\lambda \gtrapprox 0.55$.

Finally, we discuss a small minimum in $\partial^2 E/\partial \lambda^2$ for $\Gamma>0$ and $K<0$, which is expressed by the open square in Fig. \ref{fig5}(a).  
When the small minimum is caused by a crossover, QSL in Refs. \onlinecite{Yamaji2018-1} and \onlinecite{Yamaji2018-2} adiabatically connects to the $|t_x\rangle$-dimer state. If the small minimum would be caused by a phase transition, QSL does not connect with the  $|t_x\rangle$-dimer state.  
A future study is needed to obtain a definite conclusion.

\section{\label{sec:level4} Stability of the Kekul\'{e} dimerized state}
\begin{figure*}[htb]
\begin{center}
\includegraphics[width=0.19\hsize]{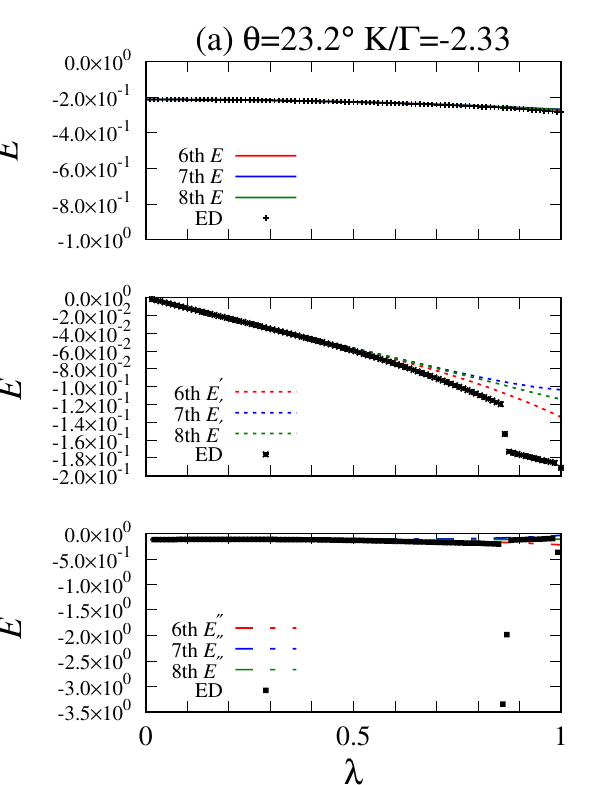}
\includegraphics[width=0.19\hsize]{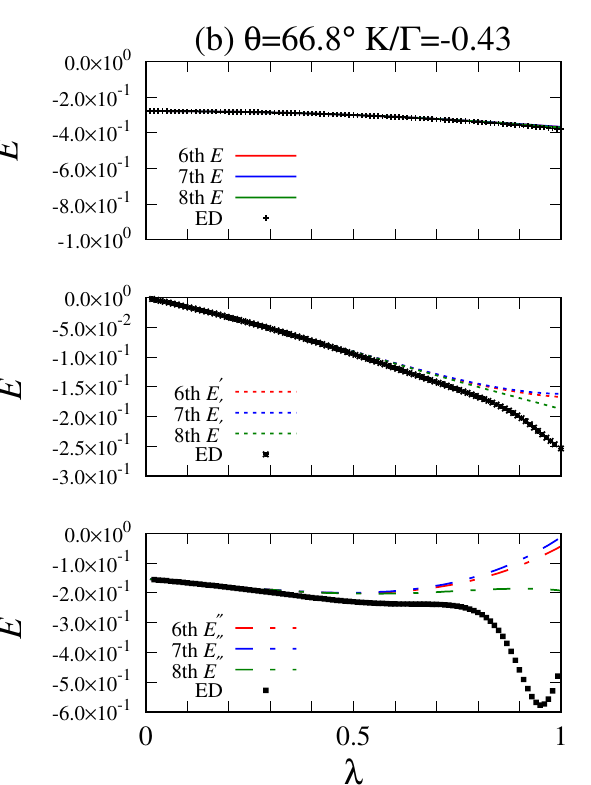}
\includegraphics[width=0.19\hsize]{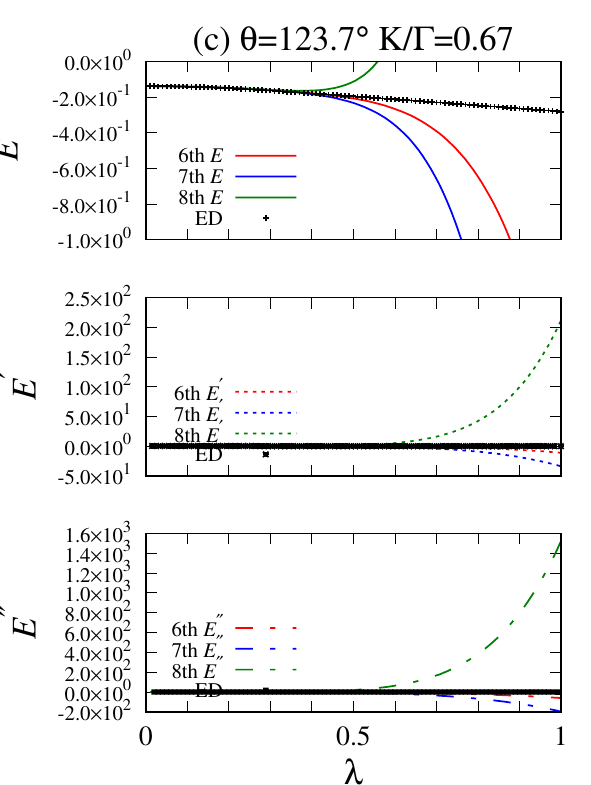}
\includegraphics[width=0.19\hsize]{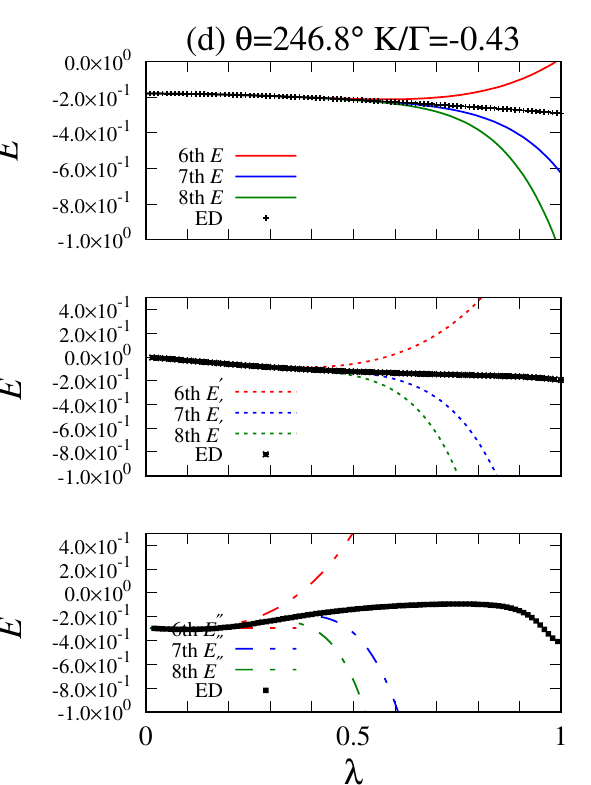}
\includegraphics[width=0.19\hsize]{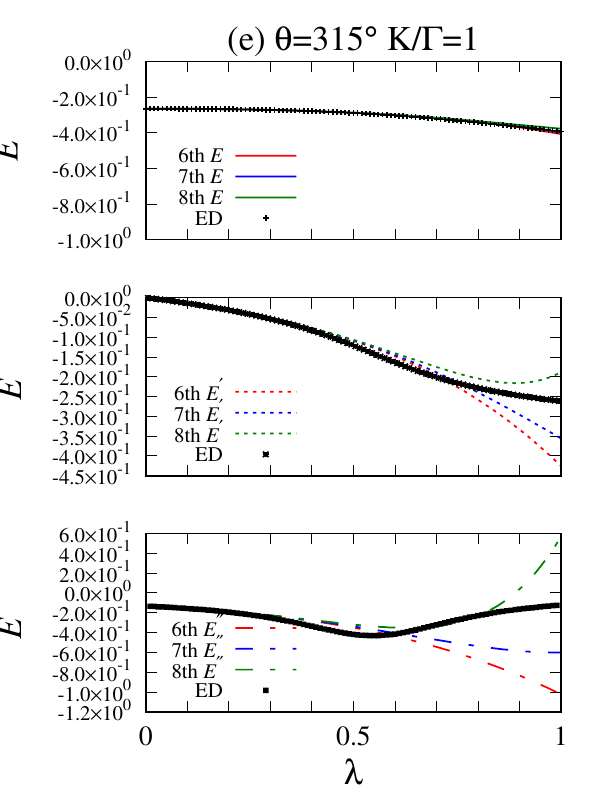}
\hspace{0pc}
\vspace{0pc}
\caption{(Color online) \label{fig7} The typical behavior of $E$, $\partial E/\partial \lambda$, and $\partial^2 E/\partial \lambda^2$ obtained with dimer series expansions performed in Sec. \ref{sec:level4} up to sixth, seventh, and eighth orders and with 24-site ED.  $\theta={\rm(a)}\; 23.2^{\circ}$, ${\rm(b)}\; 66.8^{\circ}$,  ${\rm(c)}\; 123.7^{\circ}$, ${\rm(d)}\; 246.8^{\circ}$, and ${\rm(e)}\; 315^{\circ}$. 
}
\end{center}
\end{figure*}
Our calculations have shown that in some regions, a phase transition takes place at $\lambda=1$ where $C_{3v}$ symmetry is recovered.
This stimulates us to investigate a stable ordered state that has discrete symmetry at $\lambda=1$. 
For this purpose, we adopt the ${\mathcal T}_6$ transformation \cite{JChaloupkaPRB} to Hamiltonian (1). The ${\mathcal T}_6$ transformation transforms Hamiltonian (1) into an extended $XXZ$ Hamiltonian with a Kekul\'e structure shown in Fig. \ref{fig1}(d). Note that the transformed Hamiltonian still keeps $C_{3v}$ symmetry. 
As shown in Fig. \ref{fig1}(d), we make the initial dimers form a Kekul\'e structure. We then apply dimer series expansions to the transformed Hamiltonian by treating the interactions on the mapped $X$ and $Y$ bonds as perturbation.


Figures \ref{fig7}(a)--\ref{fig7}(e) show the typical $\lambda$ dependences of $E$, $\partial E/\partial \lambda$, and $\partial^2 E/\partial \lambda^2$. We calculate these quantities with 24-site ED. Because the initial dimer placement differs from that performed in Sec. \ref{sec:level3}, the $\lambda$ dependences of $E$, $\partial E/\partial \lambda$, and $\partial^2 E/\partial \lambda^2$ differ from those shown in Sec. \ref{sec:level3}.  
For $\theta=23.2^{\circ} \; (K/\Gamma=-2.33 \; {\rm at} \; \Gamma<0)$ [(a)] and $66.8^{\circ} \;(K/\Gamma=-0.43 \; {\rm at} \; \Gamma<0)$ [(b)], the ground-state energies up to sixth, seventh, and eighth orders converge at $0 \leq \lambda \leq 1$. They agree with the ground-state energy obtained with ED. 
For $\theta=23.2^{\circ}$, $\partial E/\partial \lambda$ obtained with ED shows discontinuity at $\lambda \approx 0.86$, while $\partial E/\partial \lambda$ up to sixth, seventh, and eighth orders begin to deviate at $\lambda \approx 0.6$ with an increase in $\lambda$. These results mean that the Kekul\'{e} dimerized state undergoes a first-order phase transition at $\lambda \approx 0.86$ and becomes unstable at the $\lambda=1$ isotropic point. 
For $\theta=66.8^{\circ}$, $\partial^2 E/\partial \lambda^2$ shows a minimum at $\lambda \approx 0.96$, which indicates that the Kekul\'{e} dimerized state becomes unstable at $\lambda \gtrapprox 0.96$.   
For $\theta=123.7^{\circ} \;(K/\Gamma=0.67 \; {\rm at} \; \Gamma>0)$ [(c)] and $246.8^{\circ} \;(K/\Gamma=-0.43 \; {\rm at} \; \Gamma>0)$ [(d)], the ground-state energies up to sixth, seventh, and eighth orders deviate at $\lambda \approx 0.4$ and $\approx 0.6$, respectively, and show divergence-like behavior with an increase in $\lambda$. The results indicate that the Kekul\'{e} dimerized state becomes unstable before the system approaches the isotropic point. 
For $\theta=315^{\circ} \;(K/\Gamma=1 \; {\rm at} \; \Gamma<0)$ [(e)], the ground-state energies up to sixth, seventh, and eighth orders converge and agree with that obtained with ED at $0 \leq \lambda \leq 1$. On the other hand, $\partial^2 E/\partial \lambda^2$ obtained with ED shows a small minimum at $\lambda \approx 0.55$, and $\partial^2 E/\partial \lambda^2$ obtained with the present dimer-series expansions begin to deviate there with an increase in $\lambda$. The results suggest that the Kekul\'{e} dimerized state becomes unstable at $\lambda \gtrapprox 0.55$.

Dimer series expansions and ED performed in this section argue that the Kekul\'{e} dimerized state is unstable at $\lambda=1$ isotropic points.

\section{\label{sec:level5} Summary}
We have investigated the ground-state phase diagram of the $K-\Gamma$ model on a honeycomb lattice using dimer series expansions and ED. Starting from the initial dimers placed on a specific bond, we have strengthened the interactions between the nearest-neighbor dimers and investigate the stability of the dimer state. The results have been summarized in the phase diagrams. 
We have shown that at $\Gamma>0$ and $K<0$, the $|t_{x}\rangle$-dimer state survives up to the isotropic $\lambda=1$ point where the phase transition occurs or undergoes a first-order phase transition close to the isotropic point. We have proposed two scenarios for the connection of QSL \cite{Yamaji2018-1,Yamaji2018-2} to the $|t_x\rangle$-dimer state. 
At $\Gamma>0$ and $K>0$, the $|t_{x}\rangle$-dimer state undergoes a phase transition to the $120^{\circ}$ magnetically ordered state with an increase in $\lambda$. When $K/\Gamma$ is increased at $\Gamma>0$ and $K>0$, the critical value of $\lambda$ decreases, and the $|t_{x}\rangle$-dimer region is reduced. 
At $\Gamma<0$ and $K<0$, the $|t_{y}\rangle$-dimer state undergoes a phase transition to a ferromagnetically ordered state at $\lambda \approx 0.6$ with an increase in $\lambda$. When $K/|\Gamma|$ is increased at $\Gamma<0$ and $K>0$, the lower critical value of $\lambda$ decreases, and the $|t_{y}\rangle$-dimer region is reduced. 
We have also shown that the Kekul\'{e} dimerized state is unstable in the isotropic $K-\Gamma$ model.

\begin{acknowledgments}
This work was supported by the CDMSI, CBSM2, and JSPS KAKENHI Grants No. 16K17751 and No. 19K03721. 
We are also grateful for the numerical resources at the ISSP Supercomputer Center at the University of Tokyo  
and the Research Center for Nano-Micro Structure Science and Engineering at University of Hyogo.
\end{acknowledgments}

\end{document}